\documentclass[twocolumn]{autart}    

\usepackage{graphicx}          
 \usepackage{cite}
\usepackage{amsmath,amssymb,amsfonts}
\usepackage{amssymb}
\usepackage{algorithmic}
\usepackage{graphicx}
\usepackage{textcomp}
\usepackage{xcolor}
\usepackage{algorithm, algorithmic}
\usepackage{subfigure}
\usepackage{subfigure}
\newtheorem{assumption}{Assumption}

\newtheorem{lemma}{Lemma}

\newtheorem{theorem}{Theorem}

\newtheorem{remark}{Remark}

\begin{document}

\begin{frontmatter}

\title{A co-design method of online learning SMC law via an input-mappping strategy\thanksref{footnoteinfo}} 

\thanks[footnoteinfo]{This paper was not presented at any IFAC
meeting. Corresponding author: Dewei Li (dwli@sjtu.edu.cn).}

\author[Paestum]{Yaru Yu}\ead{ yaruyu123@sjtu.edu.cn},    
\author[Paestum]{Dewei Li}\ead{ dwli@sjtu.edu.cn},               
\author[Rome]{Dongya Zhao}\ead{dyzhao@upc.edu.cn},  
\author[Paestum]{Yugeng Xi}\ead{ygxi@sjtu.edu.cn}

\address[Paestum]{Department of Automation, Shanghai Jiao Tong University,
Key Laboratory of System Control and Information Processing, Ministry of Education of China, Shanghai, China, 200240 }  
\address[Rome]{College of  New Energy, China University of Petroleum (East China), Qingdao, China, 266580}             

\begin{keyword}                           
Sliding mode control, input-mapping, co-design.              
\end{keyword}                             

\begin{abstract}                          
The research on sliding mode control strategy is generally based on the robust approach. The larger parameter space consideration will inevitably sacrifice part of the performance. Recently, the data-driven sliding mode control method attracts much attention and shows excellent benefits in the fact that data is introduced to compensate the controller. Nevertheless, most of the research on data-driven sliding mode control relied on identification techniques, which limits its online applications due to the special requirements of the data. In this paper, an input-mapping technique is inserted into the design framework of sliding mode control to compensate for the influence generated by the unknown dynamic of the system. The major novelty of the proposed input-mapping sliding mode control strategy lies in that the sliding mode surface and the sliding mode controller are co-designed through online learning from historical input-output data to minimize an objective function. Then, the convergence rate of the system is improved significantly based on the method designed in this work. Finally, some simulations are provided to show the effectiveness and superiority of the proposed methods.
\end{abstract}

\end{frontmatter}

\section{Introduction}
It is well known that sliding mode control (SMC) is a typical nonlinear control method and it shows strong robustness to uncertain systems \cite{Y2009FSS,DJCNP2012ATO,SXZ2012ATO,WPSL2010TAC}. So far, abundant theoretical and applied achievements have been presented due to the characteristics of simplicity and flexibility of SMC strategy \cite{ABX2018ATO, CNS2020TAC, HSHZ2010TPE, HJCLH2013TIE}.

With the development and application of computer technology, considerable attention has been taken into account for SMC of discrete-time systems. The most noteworthy one is that the system trajectories of the discrete-time systems move in the vicinity of the ideal sliding mode surface. The region where the system state trajectory is ultimately maintained is the so-called quasi-sliding mode band (QSMB) \cite{HYXCC2019TM, MWX2017TIE}. In fact, such a chattering phenomenon is unfavorable to the implementation of the sliding mode controller. Therefore, many efforts focus on the research of eliminating or weakening chattering. Such as the reaching law method \cite{GWH2002TIE}, the boundary layer technique \cite{saghafinia2014adaptive}, and others \cite{li2013chattering}. Where the reaching law method proposed by Gao et al \cite{GWH2002TIE} has attracted wide attention due to its superiority in simplicity and robustness. Then, Qu et al proposed a disturbance dynamic compensation reaching law method in \cite{QXZ2014TIE} to compensate the sliding mode controller by introducing the superposition sum of the deviation between ideal reaching law and real reaching law. In this way, the chattering phenomenon is significantly weakened and the dynamic performance of the system is improved. Therefore, the proposed disturbance dynamic compensation method has enriched the theoretical achievements in the field of SMC. However, the robust design may lead to the loss of control performance owing to the considered large space of unknown dynamic.

For the past few decades, offline and online data is beginning to play a significant role in controller design. In  \cite{MXLZ2020TCS,EOR2018IJC,wang2016data,weng2017data,hou2021data1,hou2021data2,LY2017TSMCS}, the data-driven SMC methods were presented to compensate for the influence of unknown dynamic by identifying an approximate model according to data. It is undeniable that the usage of data indeed offsets the deficiency of SMC strategy based on robust methods. Nevertheless, the existing works for data-driven model-free adaptive control usually need to identify a pseudo-partial derivative model and the identification has special requirements on data which limits the parameters close to the actual value. This is difficult for the actual application.

Note that the model information of the system is implicated in its input-output data. Therefore, an input-mapping technique was proposed in \cite{YLXLX2020IET,HCLXZ2021TC} to online compensate for the impact of unknown dynamic by a linear combination of past data. Inspired by the input-mapping strategy, this paper develops an input-mapping sliding mode control (IMSMC) strategy for the system with unknown dynamic. Especially, the SMC law and the sliding variable are co-designed to minimize an objective function according to the past information and the superposition parameters are obtained by optimizing in the meantime. For the co-design method, the SMC law and the sliding variable are designed simultaneously to guarantee satisfactory system performance.

Compared with the robust sliding mode control method in \cite{QXZ2014TIE}, the input-mapping technique is inserted into the design framework of the SMC strategy to accelerate the convergence rate of the system trajectory by online input-output data. On the other hand, the design method proposed in this paper directly predicts the future system dynamic by historical data generated online. This avoids the need to adjust parameters by a trial-and-error method in \cite{MXLZ2020TCS} and the special requirements on the data used for model identification in \cite{EOR2018IJC}. Moreover, the stability of the addressed system is analyzed and demonstrated by the Lyapunov method. Some comparisons with existing methods are given in the simulation to show the superiority of the algorithm.

The rest of this paper is organized as follows. The system description and problem formulation are given in Section II. The specific input-mapping-based online learning SMC strategy is introduced in Section III. In section IV, two examples are presented to illustrate the effectiveness and advantages of the methods proposed in this paper. Section V shows some important conclusions.

Notations: For a matrix $A$, $A^{T}$ denote its transpose.  $\|x\|:=\sqrt{x^{T}x}$ stands for its 2-norm; $\mathbb{R}^n$ denotes the $n$-dimensional Euclidean space; $\mathbb{N}$ represents positive integers; $\mathrm{sgn}(\cdot)$  satisfies $\mathrm{sgn}(s(k))=1$ if $s(k)>0$ and $\mathrm{sgn}(s(k))=-1$ if $s(k)\leq0$ ; $\sup_{k\in\mathbb{N}}()$ is the maximum value of a parameter within $k\in\mathbb{N}$; $x:=\mathrm{col}(x_{1},\ldots, x_{n})=[x_{1}^{T},\ldots,x_{n}^{T}]^{T}$.
\section{ System description and problem formulation}
\vspace{-0.5em}
\subsection{System description}
\vspace{-0.5em}
In this paper, we consider a class of systems suffer from unknown-but-bounded unknown dynamic described by
\begin{eqnarray}
\varsigma(k+1)=[\tilde{A}+\triangle \tilde{A}]\varsigma(k)+\tilde{B}u(k),
\label{UBS-1}
\end{eqnarray}
where $\varsigma(k)\in\mathbb{R}^{n_{\zeta}}$ is the state and $u(k)\in\mathbb{R}^{n_{u}}$ is the control input at $k\in \mathbb{N}$. $\tilde{A}$, $\bigtriangleup \tilde{A}$ and $\tilde{B}$  represent the system matrix, the system unknown dynamic and the input matrix, respectively. Then, some assumptions are given
 for the subsequent derivation.
\begin{assumption}\label{Assumption-1}
The unknown dynamic $\triangle \tilde{A}$ satisfies $\triangle \tilde{A}=D\Delta E$, where $D$ and $E$ are known matrices. $\Delta$ is unknown and it satisfies $\Delta^T\Delta\leq I$.
\end{assumption}
\begin{assumption}\label{Assumption-2}
The pair $(\tilde{A}, \tilde{B})$ is controllable and the matrix $\tilde{B}$ is column full rank.
\end{assumption}

 Now, the input matrix $\tilde{B}$ can be partitioned as $\tilde{B}=[\tilde{B}^T_{1}~\tilde{B}^T_{2}]^{T}$ from the Assumption \ref{Assumption-2} and $\tilde{B}_{2}$ is invertible. Then, by implementing a coordinate transformation $x(k)=T_{c}\varsigma(k)\in \mathbb{R}^{n_{x}} $ $(n_{x}=n_{\varsigma})$ with
 \begin{eqnarray*}
T_{c}\in\mathbb{R}^{n_{x}\times n_{x}}=\left[\hspace{-0.2em}\begin{array}{cc} I_{(n_{x}-n_{u})\times(n_{x}-n_{u})}& -\tilde{B}_{1}\tilde{B}^{-1}_{2}\\ 0_{n_{u}\times (n_{x}-n_{u})}&I_{n_{u}\times n_{u}}\end{array}\right].
\label{MJS-A4}
 \end{eqnarray*}
 Then, the system (\ref{UBS-1}) can be reshaped as
 \begin{eqnarray}
x(k+1)=(A+\triangle A)x(k)+Bu(k).
\label{UBS-2}
\end{eqnarray}
Specifically, $T_{c}$ is an invertible matrix. Matrices $A$, $B$ and $\triangle A$ satisfy $A=T_{c}\tilde{A}T^{-1}_{c}$, $B=T_{c}\tilde{B}$ and $\triangle A=T_{c}\triangle \tilde{A}T^{-1}_{c}$. Then, we have (\ref{UBS-2}) is equivalent to
\allowdisplaybreaks
\begin{eqnarray}
\left[\begin{array}{cc}x_{1}(k+1)\\x_{2}(k+1)\end{array}\hspace{-0.3em}\right]=\left(\left[\begin{array}{cc}A_{11}&A_{12}\\ A_{21}&A_{22}\end{array}\right]+
\left[\begin{array}{cc}\bigtriangleup A_{11}&\bigtriangleup A_{12}\\ \bigtriangleup A_{21}&\bigtriangleup A_{22}\end{array}\right]\hspace{-0.2em}\right)\notag\\
\times\left[\begin{array}{cc}x_{1}(k)\\x_{2}(k)\end{array}\right]+\left[\begin{array}{cc} 0_{(n_{x}-n_{u})\times n_{u}}\\ B_{1}\end{array}\right]u(k),
\label{UBS-3}
\end{eqnarray}
and the unknown dynamic is supposed to be represented by
\begin{eqnarray}
\left[\begin{array}{cc}\bigtriangleup A_{11}&\bigtriangleup A_{12}\\ \bigtriangleup A_{21}&\bigtriangleup A_{22}\end{array}\right]=\bar{D}\Delta\bar{E}=\left[\begin{array}{cc}\bar{D}_{1}\\ \bar{D}_{2}\end{array}\right]\Delta[\begin{array}{cc}\bar{E}_{1}&\bar{E}_{2}\end{array}].
\label{UBS-4}
\end{eqnarray}
\begin{remark}
It can be observed from (\ref{UBS-3}) that the state component  $x_{1}(k+1)$ is independent of control input $u(k)$, and only the state $x_{2}(k+1)$ is related to $u(k)$. The system model in this form  is the so-called “regular form” as in \cite{su2018sliding} for the convenience of designing a SMC strategy.
\end{remark}
\subsection{Preliminaries}
\subsubsection{Sliding mode surface design}
In this section, an appropriate sliding mode surface is designed to guarantee the stability of the states which reach the surface.
Without loss of generality, the sliding mode surface for system (\ref{UBS-2}) is considered as
\begin{eqnarray}
s(k)=\bar{G}x(k)=[\begin{array}{cc}G & I \\ \end{array}]
x(k),
\label{UBS-5}
\end{eqnarray}
where $G\in \mathbb{R}^{n_{u}\times (n_{x}-n_{u})}$ is the sliding variable to be designed later. Then, the corresponding sliding manifold is assigned as
\begin{eqnarray}
\mathcal{\varrho}_{x}\triangleq\left\{x\in \mathbb{R}^{n_{x}}|s=\bar{G}x(k)=0\right\}.
\label{UBS-6}
\end{eqnarray}
It is indisputable that the system trajectories arrive at sliding manifold (\ref{UBS-6})
satisfies $s(k)=0$, that is, $x_{2}(k)=-Gx_{1}(k)$. Substituting $x_{2}(k)$ of the first term  in (\ref{UBS-3}) by the established relation, the sliding mode dynamic (SMD) of the considered system (\ref{UBS-3}) can be obtained as
\begin{eqnarray}
x_{1}(k+1)=A_{g}x_{1}(k),
\label{UBS-7}
\end{eqnarray}
with $A_{g}=(A_{11}+\Delta A_{11})-(A_{12}+\Delta A_{12})G$. Apparently, the sliding mode variable $G$ can be designed by  recognizing $x_{2}(k)$  as a virtual control input of $x_{1}(k)$. Then, a linear matrix  inequality condition is given in Lemma \ref{lemma-1} to guarantee the robustness of the SMD (\ref{UBS-7}) by employing quadratic stability analysis.
\begin{lemma}\label{lemma-1}
\cite{C2002ATO} For a chosen Lyapunov function $V(k)=x^{T}_{1}(k)Rx_{1}(k)$, the SMD in (\ref{UBS-7}) is asymptotically stable if there exist scalar $\gamma>0$,  matrix $R_{1}=R^{-1}$ is positive definite and $R_{g}$ is a matrix with appropriate dimension satisfy
\begin{eqnarray}
\left[
  \begin{array}{ccc}
    -R_{1} & R_{1}A^{T}_{11}-R_{g}A^{T}_{12} & R_{1}\bar{E}^{T}_{1}-R_{g}\bar{E}^{T}_{2}\\
    * & -R_{1}+\gamma\bar{D}_{1}\bar{D}^{T}_{1} & 0 \\
    * & * & -\gamma I \\
  \end{array}
\right]<0,
\label{UBS-8}
\end{eqnarray}
and then the variable $G$ is derived off-line as $G=R_{g}R_{1}^{-1}$.
\end{lemma}
Then, the disturbance compensation SMC law proposed in  \cite{QXZ2014TIE} is presented to ensure the system trajectories from arbitrary initial state will converge to a QSMB.
\subsubsection{SMC law}
In \cite{QXZ2014TIE}, a discrete-time reaching law with a disturbance compensator is provided by utilizing the deviation between the real reaching law and the ideal reaching law. The reaching law is shown below:
\begin{eqnarray}
s(k+1)-s(k)=-\mu T s(k)-\xi T\mathrm{sgn}s(k)+\varpi(k)\notag\\
-\sum^{k}_{i=2}\{s(i)-[(1-\mu T)s(k-i)-\xi T\mathrm{sgn}s(k)]\},
\label{UBS-9}
\end{eqnarray}
and
\allowdisplaybreaks
\begin{eqnarray*}
&&\varpi(k)\hspace{-0.2em}=\hspace{-0.2em}\bar{G}\Delta Ax(k),\\
&&\varpi(k-1)\hspace{-0.2em}=\hspace{-0.2em}\sum^{k}_{i=2}\{s(i)\hspace{-0.2em}-\hspace{-0.2em}[(1-\mu T)s(k-i)-\xi T\mathrm{sgn}s(k)]\},
\end{eqnarray*}
 where $\mu$ is a converging parameter and $\xi$ is a switching parameter. $T$ is the sampling period, and $1>1-\mu T>0$. Assume that the gradient of parameter perturbation within the system is bounded by $\|\varpi(k)-\varpi(k-1)\|<\delta$. Then, the following result is derived to ensure the system trajectories of system (\ref{UBS-3}) can be driven onto the designed sliding surface in finite time and stay on there all the succeeding time.
\begin{lemma}\label{lemma-2}
\cite{QXZ2014TIE}
On the basis of the sliding variable $G$ designed by Lemma \ref{lemma-1}, the trajectories of the considered system (\ref{UBS-2}) can be driven onto a QSMB in finite time by the following controller:
\allowdisplaybreaks
\begin{eqnarray}
u(k)&\hspace{-0.2em}=&\hspace{-0.2em}-(\bar{G}B)^{-1}\left[\mu Ts(k)+\xi T\mathrm{sgn}s(k)+\bar{G}Ax(k)-s(k)\right.\notag\\
&\hspace{-0.2em}+&\hspace{-0.2em}\sum^{k}_{i=2}\{s(i)-[(1-\mu T)s(k-i)-\xi T\mathrm{sgn}s(k)]\}].
\label{UBS-10}
\end{eqnarray}
\end{lemma}

\begin{remark}
The SMC strategy based on Lemmas \ref{lemma-1}-\ref{lemma-2} is designed with the idea of robustness. Hence, the control performance sacrifices are inevitably by the derived sliding mode controller due to the considered space of unknown dynamic being relatively large. With the fact that the unknown dynamic of the addressed system show the same characteristic as in the past, it is a natural idea to collect historical information to compensate for the designed SMC strategy and for improving the convergence rate of the system trajectory.
\end{remark}

\section{Input-mapping-based co-deign SMC strategy}
We now aim at synthesizing a valid input-mapping-based online learning SMC law to drive the state trajectories of the closed-loop system in (\ref{UBS-3}) to $\varrho_{x}$. Particularly, the sliding mode surface in (\ref{UBS-5}) is continuously adjusted by online learning from the past behavior of the system until it enters the QSMB. That is,  the sliding mode variable $G(k)$ and sliding mode controller are co-designed by minimizing a given objective function, which is different from the methods in \cite{su2018sliding,zhang2018discrete,wang2021observer}.

In order to form an input-mapping-based online learning SMC strategy for the addressed system, the following section is given for some preliminary presentations.
\subsection{Input-mapping technique}
As for the system (\ref{UBS-2}), we can get the following relations
\begin{eqnarray}
x(k)=(A+\triangle A)x(k-1)+Bu(k-1).
\end{eqnarray}\label{UBS-22}
Similar to \cite{YLXLX2020IET,HCLXZ2021TC}, at time $k$, the unknown dynamic $\triangle Ax(k-1)$ of considered system can be calculated by
\begin{eqnarray}
\triangle Ax(k-1)=x(k)-Ax(k-1)-Bu(k-1),
\end{eqnarray}\label{UBS-22-1}
where $x(k)$, $x(k-1)$ and $u(k-1)$ are known information to exactly derive unknown dynamic $\triangle Ax(k-1)$. Similarly, at time $k$, we have the historical relations
\begin{eqnarray}
x(k-i+1)\hspace{-0.2em}=\hspace{-0.2em}(A\hspace{-0.2em}+\hspace{-0.2em}\triangle A)x(k-i)\hspace{-0.2em}+\hspace{-0.2em}Bu(k-i),~ i\in\mathbb{N},\label{UBS-22-2}
\end{eqnarray}
where $\Delta Ax(k-i)$ can be uniquely determined by a set of known historical data $\{x(k-i+1), x(k-i), u(k-i)\}$  although the item $\Delta A$ is unknown.

 On the other hand, it is a fact that for a vector in $n$-dimensional linear space, it can be uniquely linearly represented by $n$ linearly independent basis vectors. In that way, the system state $x(k)\in \mathbb{R}^{n}$ can be uniquely linearly represented by $n$ linearly independent basis vectors. Thus, if the historical system states of $N$ moments near the current moment are collected as the basis vectors of the corresponding space, the system states $x(k)$ can be decomposed as
\begin{eqnarray}
x(k)=\sum^{N}_{i=1}l_{i}x(k-i)+\delta_{x}(k), \label{UBS-22-5}
\end{eqnarray}
where $l_{i}$ is the unknown combination parameter, $x(k-i)$ are the known system states at time $k$, and $\delta_{x}(k)$ is the residual item to express $x(k)$ because the linear independence of $N$ historical data may not be guaranteed online. Analogously, the control input $u(k)$ can be decomposed as
\begin{eqnarray}
u(k)=\sum^{N}_{i=1}l_{i}u(k-i)+\delta_{u}(k),\label{UBS-11}
\end{eqnarray}
where $l_{i}$ is the same as in (\ref{UBS-22-5}), $u(k-i)$ are the known control inputs at time $k$, and $\delta_{u}(k)$ is also a residual item to represent $u(k)$. Moreover, based on the collected $N$ historical data $x(k-i)$ and $u(k-i)$ near the current moment, (\ref{UBS-22-2}) can be converted to
\begin{eqnarray}
&&\sum^{N}_{i=1}l_{i}x(k-i+1)\nonumber\\
&&=(A\hspace{-0.2em}+\hspace{-0.2em}\triangle A)\sum^{N}_{i=1}l_{i}x(k-i)+B\sum^{N}_{i=1}l_{i}u(k-i),\label{UBS-22-3}
\end{eqnarray}
by the homogeneity property and superposition principles of a linear system. Based on (\ref{UBS-22-3}), the system model in (\ref{UBS-2}) can be reshaped by substituting the expressions  (\ref{UBS-22-5}) and (\ref{UBS-11}) as
\begin{eqnarray}
x(k&+&1)=(A+\Delta A)[\sum^{N}_{i=1}l_{i}x(k-i)+\delta_{x}(k)]\nonumber\\
&&+B[\sum^{N}_{i=1}l_{i}u(k-i)+\delta_{u}(k)]\nonumber\\
&=&(A+\Delta A)\sum^{N}_{i=1}l_{i}x(k-i)+B\sum^{N}_{i=1}l_{i}u(k-i)\nonumber\\
&&+(A+\Delta A)\delta_{x}(k)+B\delta_{u}(k)\nonumber\\
&=&\sum^{N}_{i=1}l_{i}x(k-i+1)\hspace{-0.2em}+\hspace{-0.2em}(A+\Delta A)\delta_{x}(k)+B\delta_{u}(k).
\label{UBS-13-2}
\end{eqnarray}
For simplifying the subsequent derivation, the relations (\ref{UBS-22-5}), (\ref{UBS-11}) and (\ref{UBS-13-2}) are rewritten as matrix form as follows
\begin{eqnarray}
&&x(k)=\mathcal{X}(k-1)L(k)+\delta_{x}(k),\label{UBS-13-1}\\
&&u(k)=\mathcal{U}(k-1)L(k)+\delta_{u}(k),\label{UBS-13}\\
&&x(k+1)\hspace{-0.3em}=\hspace{-0.3em}\mathcal{X}(k)L(k)+(A+\Delta A)\delta_{x}(k)+B\delta_{u}(k),
\label{UBS-14}
\end{eqnarray}
where
\begin{eqnarray}
\mathcal{X}(k)&\hspace{-0.2em}=&\hspace{-0.2em}\left[x(k), x(k-1),\ldots, x(k-N+1)\right]\in \mathbb{R}^{n_{x}\times N},\nonumber\\
\mathcal{X}(k-1)&\hspace{-0.2em}=&\hspace{-0.2em}\left[x(k-1), x(k-2),\ldots, x(k-N)\right]\in \mathbb{R}^{n_{x}\times N},\nonumber\\
\mathcal{U}(k-1)&\hspace{-0.2em}=&\hspace{-0.2em}\left[u(k-1), u(k-2),\ldots, u(k-N)\right]\in \mathbb{R}^{n_{u}\times N},\nonumber\\
L(k)&\hspace{-0.2em}=&\hspace{-0.2em}\left[l_{1}, l_{2}, \ldots, l_{N}\right]^{T}\in \mathbb{R}^{N}.\label{UBS-15-1}
\end{eqnarray}

From (\ref{UBS-14}), we have vector $\mathcal{X}(k)$ is fully known historical data. The vector variable $L(k)$ and the residual items $\delta_{x}(k)$ and $\delta_{u}(k)$ are unknown terms need to solve. At time $k$, the control problem of designing $u(k)$ is transformed into a problem of solving $L(k)$ and $\delta_{u}(k)$. Specially, the unknown dynamic is just related to $\delta_{x}(k)$, which is only a part of system state $x(k)$. Therefore, the influence caused by part of the unknown dynamic of the system can be compensated by historical information, which can effectively improve the system's performance.

Furthermore, a data table needs to be created to store historical information. The data table is updated at every moment, and the new data is stored in the table and the data far away from the current moment is removed. Then, the information stored in the data table is the basis for online optimization calculations.

\subsection{Co-design IMSMC law}
  In this section, a developed input-mapping-based online learning SMC law will be constructed by a co-designed method for improving the system performance. The sliding mode surface is online adjusted in terms of historical data to ensure that the system state can reach the sliding surface faster. Firstly, the sliding mode surface is rewritten as
  \begin{eqnarray}
s(k)=\bar{G}(k)x(k)=[\begin{array}{cc}G(k) & I \\ \end{array}]
x(k),
\label{UBS-17-2}
\end{eqnarray}
where the sliding variable $G(k)$ is time-dependent now. The SMD is regained as
\begin{eqnarray}
\left\{\begin{array}{ll}x_{1}(k+1)=\bar{A}_{g}x_{1}(k)\\\bar{A}_{g}=(A_{11}+\Delta A_{11})+(A_{12}+\Delta A_{12})G(k)\end{array}\right..
\label{UBS-18-1}
\end{eqnarray}
Then, the reaching law is given as follows
\begin{eqnarray}
s(k+1)&-&s(k)=-\bar{\mu}(k) T s(k)-\bar{\xi} T\mathrm{sgn}s(k)
+\varpi_{1}(k)\notag\\
&&\hspace{-2.5em}-\sum^{k}_{i=2}\{s(i)\hspace{-0.2em}-\hspace{-0.2em}[(1\hspace{-0.2em}-\hspace{-0.2em}\bar{\mu}(k) T)s(k\hspace{-0.2em}-\hspace{-0.2em}i)\hspace{-0.2em}-\hspace{-0.2em}\bar{\xi} T\mathrm{sgn}s(k)]\},
\label{UBS-18}
\end{eqnarray}
with
\begin{eqnarray*}
\left\{\begin{array}{llll}s(k+1)=\bar{G}(k+1)x(k+1),\\
s(k)=\bar{G}(k)x(k),\\
\varpi_{1}(k)=\bar{G}(k+1)\Delta A\delta_{x}(k).\end{array}\right.
\end{eqnarray*}
Similar to the derivation of \cite{QXZ2014TIE}, we have $\varpi_{1}(k-1)=\sum^{k}_{i=2}\{s(i)\hspace{-0.2em}-\hspace{-0.2em}[(1\hspace{-0.2em}-\hspace{-0.2em}\bar{\mu}(k) T)s(k\hspace{-0.2em}-\hspace{-0.2em}i)\hspace{-0.2em}-\hspace{-0.2em}\bar{\xi} T\mathrm{sgn}s(k)]\}$.
The convergence parameter $\bar{\mu}(k)$ satisfy $1>1-\bar{\mu}(k)T>0$. $\bar{\xi}$ is the switching parameter.
\begin{remark}
As the sliding mode variable is a determined value solved in advance, and then the convergence parameter for the conventional way is a given constant that satisfy certain conditions. However, the sliding mode surface for co-design IMSMC strategy is online adjusted by leaning historical behavior. Thus, the convergence parameter $\bar{\mu}(k)$ for co-design IMSMC strategy (sliding variable is time-dependent) is assigned to time-varying due to the reachability of the system is difficult to guarantee by a constant convergence parameter.
\end{remark}

On the basis of sliding mode surface (\ref{UBS-5}), model (\ref{UBS-14}) and reaching law (\ref{UBS-18}), the input-mapping-based SMC law is established as follows
\begin{eqnarray}
\delta_{u}(k)&\hspace{-0.15em}=&\hspace{-0.15em}-(\bar{G}(k+1)B)^{-1}\left[\bar{\mu}(k) T s(k)+\bar{\xi} T \mathrm{sgn}s(k)\hspace{-0.2em}-\hspace{-0.2em}s(k)\right.\notag\\
&\hspace{-0.15em}+&\hspace{-0.15em}\bar{G}(k+1)\mathcal{X}(k)L(k)+\bar{G}(k+1)A\delta_{x}(k)\notag\\
&\hspace{-0.15em}+&\hspace{-0.15em}\sum^{k}_{i=2}\left\{s(i)\hspace{-0.3em}-\hspace{-0.3em}[(1\hspace{-0.3em}-\hspace{-0.3em}\bar{\mu}(k) T)s(k\hspace{-0.2em}-\hspace{-0.2em}i)\hspace{-0.2em}-\hspace{-0.2em}\bar{\xi} T\mathrm{sgn}s(k)]\right\}],
\label{UBS-15}
\end{eqnarray}
and $\delta_{x}(k)=x(k)-\mathcal{X}(k-1)L(k)$. Then, $u(k)$ can be obtained by (\ref{UBS-13}), that is
\begin{eqnarray}
u(k)&\hspace{-0.1em}=&\hspace{-0.1em}\mathcal{U}(k-1)L(k)\hspace{-0.2em}-(\bar{G}(k+1)B)^{-1}\left[\bar{\mu}(k) T s(k)-s(k)\right.\notag\\
&\hspace{-0.1em}+&\hspace{-0.1em}\bar{\xi} T \mathrm{sgn}s(k)\hspace{-0.25em}+\hspace{-0.25em}\bar{G}(k\hspace{-0.25em}+\hspace{-0.25em}1)\mathcal{X}(k)L(k)\hspace{-0.3em}+\hspace{-0.3em}\bar{G}(k\hspace{-0.3em}+\hspace{-0.3em}1)A\delta_{x}(k)\notag\\
&\hspace{-0.1em}+&\hspace{-0.1em}\sum^{k}_{i=2}\{s(i)\hspace{-0.3em}-\hspace{-0.3em}[(1\hspace{-0.3em}-\hspace{-0.3em}\bar{\mu}(k) T)s(k\hspace{-0.2em}-\hspace{-0.2em}i)\hspace{-0.2em}-\hspace{-0.2em}\bar{\xi} T\mathrm{sgn}s(k)]\}],
\label{UBS-16}
\end{eqnarray}
where $L(k)$ and $G(k+1)$ are parameters need to be designed.
\begin{remark}
The control law in (\ref{UBS-16}) is equivalent to
\begin{eqnarray*}
u(k)&=&-(\bar{G}(k+1)B)^{-1}\left[\bar{\mu} Ts(k)+\bar{\xi} T\mathrm{sgn}s(k)-s(k)\right.\notag\\
&+&\bar{G}(k+1)Ax(k)+\bar{G}(k)\Delta A\delta_{x}(k-1)\notag\\
&+&\left.\bar{G}(k+1)\Delta A\mathcal{X}(k-1)L(k)\right],
\end{eqnarray*}
by implementing a transformation. In contrast to the control law (\ref{UBS-10}), the co-design IMSMC law in (\ref{UBS-16}) involves the unknown dynamic of the past. Thus, the sliding mode controller can be compensated by online learning from the past behavior. Moreover, a certain improvement of system performance can be obtained by compensating for the conservative processing of unknown dynamic in the robust method in \cite{QXZ2014TIE}.
\end{remark}

Consider $\bar{\mu}(k)T$ as a whole, i.e. $\bar{\mu}_{0}(k)\triangleq\bar{\mu}(k)T$. Note that the optimal target for system ($\ref{UBS-2}$) is to control the system states of next time, that is, $x(k+1)$ reached stabilization quickly. Then, the optimal problem can be naturally designed as follows
\begin{eqnarray}
\vspace{-1em}
J_{1}&=&\min_{\phi_{0}(k)}\parallel x(k+1)\parallel^{2},\notag\\
&=&\min_{\phi_{0}(k)}\parallel W(k)L(k)\hspace{-0.2em}+\hspace{-0.2em}\Xi(k)\hspace{-0.2em}-\hspace{-0.2em}B[\bar{G}(k+1)B]^{-1}\notag\\
&&\hspace{-0.6em}\times [\bar{\mu}(k) T s(k)\hspace{-0.2em}+\hspace{-0.2em}\bar{\xi} T \mathrm{sgn}s(k)\hspace{-0.2em}-\hspace{-0.2em}s(k)\hspace{-0.2em}+\hspace{-0.2em}\varpi_{1}(k\hspace{-0.2em}-\hspace{-0.2em}1)]\parallel^{2},
\label{UBS-19}
\end{eqnarray}
with
\begin{eqnarray*}
\phi_{0}(k)&=&\{L(k),G(k+1),\bar{\mu}(k)\},\\
W(k)&=&\mathcal{X}(k)-B[\bar{G}(k+1)B]^{-1}\bar{G}(k+1)\mathcal{X}(k),\\
\Xi(k)&=&[A+\Delta A-B[\bar{G}(k+1)B]^{-1}\bar{G}(k+1)A]\delta_{x}(k).
\end{eqnarray*}
Note that (\ref{UBS-19}) contains unknown dynamic $\Delta A$ in $\Xi(k)$, and in view of $\delta_{x}(k)$ is undecided. Then, the $\Xi(k)$ in (\ref{UBS-19}) can be removed and the objective function is refreshed as
\begin{eqnarray}
J&=&\min_{\phi_{0}(k)}\parallel W(k)L(k)-B(\bar{G}(k+1)B)^{-1}[\bar{\mu}_{0}(k) s(k)\notag\\
&&+\bar{\xi}T\mathrm{sgn}s(k)-s(k)+\hspace{-0.2em}\varpi_{1}(k\hspace{-0.2em}-\hspace{-0.2em}1)]\parallel^{2}.
\label{UBS-20}
\end{eqnarray}

 Our purpose now is to minimize the objective $J$ based on the optimization variables $L(k)$, $G(k+1)$ and $\bar{\mu}_{0}(k)$. Then, taking the derivative with respect to $L(k)$, $G(k+1)$ and $\bar{\mu}_{0}(k)$, the analytical solution of the optimal problem is obtained as a set of nonlinear matrix equations as the Box. I.
\begin{figure*}[!t]
\begin{center}
\normalsize
\newlength{\mylength}
\setlength{\fboxsep}{1.2pt}
\setlength{\mylength}{1.01\linewidth}
\addtolength{\mylength}{-2\fboxsep}
\fbox{%
\parbox{\mylength}{
\setlength{\abovedisplayskip}{-3pt}
\setlength{\belowdisplayskip}{-3pt}
\begin{eqnarray}
\left\{\begin{array}{l}\frac{\partial J}{\partial L(k)}\hspace{-0.2em}=\hspace{-0.2em}2W^{T}(k)\left\{W(k)L(k)-B(\bar{G}(k+1)B)^{-1}[\bar{\mu}_{0}(k)s(k)\hspace{-0.2em}+\hspace{-0.2em}\bar{\xi} T\mathrm{sgn}s(k)\hspace{-0.2em}-\hspace{-0.2em}s(k)\hspace{-0.2em}+\hspace{-0.2em}\varpi_{1}(k\hspace{-0.2em}-\hspace{-0.2em}1)]\right\}=0,\\
\frac{\partial J}{\partial G(k+1)}\hspace{-0.2em}=2\mathcal{X}_{1}(k)L(k)L^{T}(k)\mathcal{X}^{T}_{1}(k)G^{T}(k+1)+2\mathcal{X}_{1}(k)L(k)[\bar{\mu}_{0}(k) s(k)+\bar{\xi} T\mathrm{sgn}s(k)\hspace{-0.2em}-\hspace{-0.2em}s(k)\hspace{-0.2em}+\hspace{-0.2em}\varpi_{1}(k\hspace{-0.2em}-\hspace{-0.2em}1)]^{T}=0,\\
\frac{\partial J}{\partial \bar{\mu}_{0}(k)}\hspace{-0.2em}=\hspace{-0.2em}2[-B(\bar{G}(k+1)B)^{-1}s(k)]^{T}
\left\{W(k)L(k)\hspace{-0.2em}-\hspace{-0.2em}B(\bar{G}(k\hspace{-0.2em}+\hspace{-0.2em}1)B)^{-1}[\bar{\mu}_{0}(k) s(k)\hspace{-0.2em}+\hspace{-0.2em}\bar{\xi} T\mathrm{sgn}s(k)\hspace{-0.2em}-\hspace{-0.2em}s(k)+\varpi_{1}(k\hspace{-0.2em}-\hspace{-0.2em}1)]\right\}=0,
                            \end{array}\right.
\label{UBS-21}
\end{eqnarray}
with $\mathcal{X}(k)=\left[\begin{array}{l}\mathcal{X}_{1}(k)_{(n_{x}-n_{u})\times N}\\ \mathcal{X}_{2}(k)_{n_{u}\times N}\end{array}\right].$
}}
\end{center}
\vspace{3pt}
\centerline{Box I. }\label{box1}
\end{figure*}
Next, the co-design IMSMC law constructed in (\ref{UBS-16}) can be implemented directly by means of the solution of combination coefficient $L(k)$, sliding variable $G(k+1)$  and convergence parameter $\bar{\mu}_{0}(k)$ in (\ref{UBS-21}).
\begin{remark}
The sliding variable $G(k+1)$ in sliding mode surface (\ref{UBS-17-2}) and the parameters $\mu_0(k)$ and $L(k)$ in sliding mode controller (\ref{UBS-16}) are obtained simultaneously by optimizing. Such a co-design method is benefit to accelerate the convergence of the system trajectory because of the sliding surface is adjusted to reduce the time for the system state to reach it.
\end{remark}

Moreover, the following theorem is presented for further analysis of the proposed co-design IMSMC law.
\begin{theorem}
Assume that the gradient of parameter perturbation within the system is bounded by $\|\varpi_{1}(k)-\varpi_{1}(k-1)\|<\bar{\delta}$. For the sliding manifold in (\ref{UBS-17-2}) and the parameters $L(k)$, $G(k+1)$ and $\bar{\mu}_{0}(k)$ in (\ref{UBS-16})  derived by solving nonlinear matrix equations (\ref{UBS-21}), we come to some conclusions as follows

(1): The system trajectories of (\ref{UBS-2}) can be driven onto a quasi-sliding mode band and maintain in the QSMB all the subsequent time by IMSMC law (\ref{UBS-16}). The QSMB is bounded by
\begin{eqnarray}\label{UBS-25}
\Upsilon=\{s(k)|\parallel s(k)\parallel\leq\Omega\},
\end{eqnarray}
with
\allowdisplaybreaks
\begin{eqnarray*}
&&\Omega\hspace{-0.2em}=\hspace{-0.2em}\sup_{k\in{\mathbb{N}}}\sqrt{\bar{\Phi}^{2}(k)\hspace{-0.2em}+\hspace{-0.2em}2\eta(k)\bar{\Phi}(k)},\notag\\
&&\eta(k)\hspace{-0.2em}=\hspace{-0.2em}(1\hspace{-0.2em}-\hspace{-0.2em}\bar{\mu}_{0}(k))(\sqrt{n_{u}}\bar{\xi} T\hspace{-0.2em}-\hspace{-0.2em}\bar{\delta}),\\
&&\bar{\Phi}(k)\hspace{-0.2em}=\hspace{-0.2em}\frac{(\sqrt{n_{u}}\bar{\xi} T+\bar{\delta})^2}{2(1-\bar{\mu}_{0}(k))(\sqrt{n_{u}}\bar{\xi} T-\bar{\delta})}.
\end{eqnarray*}

(2): Let sliding variable $G(k+1)$ keeps the value calculated at the previous time unchanged when the sliding mode surface
enters the QSMB (\ref{UBS-25}). Then the stability of the SMD $(\ref{UBS-18-1})$ can be guaranteed by the solved $G(k+1)$ in $(\ref{UBS-21})$.

\end{theorem}

\textbf{Proof}
(1)
For the co-designed IMSMC strategy in (\ref{UBS-16}),  the sliding incremental can be obtained  as
\allowdisplaybreaks
\begin{eqnarray*}
\left\{\begin{array}{l}s(k+1)\hspace{-0.1em}-\hspace{-0.1em}s(k)\hspace{-0.1em}=\hspace{-0.1em}-\bar{\mu}_{0}(k)s(k)-\bar{\xi} T\mathrm{sgn}s(k)+\Delta\varpi_{\delta}(k),\\
\Delta\varpi_{\delta}=\varpi_{1}(k)-\varpi_{1}(k-1).\end{array}\right.
\end{eqnarray*}
Denote $V(k)=\frac{1}{2}s^{T}(k)s(k)$, and suppose that $\|\Delta\varpi_{\delta}\|<\bar{\delta}$. Then, we have
\allowdisplaybreaks
\begin{eqnarray}
\Delta V(k)&=&s^{T}(k)\Delta s(k)+\frac{1}{2}\Delta s^{T}(k)\Delta s(k)\notag\\
&=&\hspace{-0.2em}-\bar{\mu}_{0}(k)(1\hspace{-0.2em}-\hspace{-0.2em}\frac{1}{2}\bar{\mu}_{0}(k))\parallel s(k)\parallel^{2}\hspace{-0.2em}-\hspace{-0.2em}\sqrt{n_{u}}\bar{\xi} T\parallel s(k)\parallel\notag\\
&&+\frac{1}{2}n_{u}\bar{\xi}^{2}T^{2}+\sqrt{n_{u}}\bar{\mu}_{0}(k)\bar{\xi} T\parallel s(k)\parallel\notag\\
&&+(1-\mu_{0}(k))s^{T}(k)\Delta\varpi_{\delta}-\bar{\xi} T\mathrm{sgn}s(k)\Delta\varpi_{\delta}\notag\\
&&+\frac{1}{2}\Delta\varpi_{\delta}^{T}(k)\Delta\varpi_{\delta}\notag\\
&<&-\sqrt{n_{u}}\bar{\xi} T\parallel s(k)\parallel+\frac{1}{2}n_{u}\bar{\xi}^{2}T^{2}\notag\\
&&+\sqrt{n_{u}}\bar{\mu}_{0}(k) \bar{\xi} T\parallel s(k)\parallel\hspace{-0.2em}+\hspace{-0.2em}(1\hspace{-0.2em}-\hspace{-0.2em}\bar{\mu}_{0}(k))\parallel s(k)\parallel\bar{\delta}\notag\\
&&+\sqrt{n_{u}}\bar{\xi} T\bar{\delta}+\frac{1}{2}\bar{\delta}^{2}\notag\\
&=&-(1-\bar{\mu}_{0}(k))(\sqrt{n_{u}}\bar{\xi} T-\bar{\delta})\notag\\
&&\times\left(\parallel s(k)\parallel-\frac{(\sqrt{n_{u}}\bar{\xi} T+\bar{\delta})^2}{2(1-\bar{\mu}_{0}(k))(\sqrt{n_{u}}\bar{\xi} T-\bar{\delta})}\right)\notag\\
&\triangleq&-\eta(k)\left(\parallel s(k)\parallel-\bar{\Phi}(k)\right),
\label{UBS-27-1}
\end{eqnarray}
and $\Delta s(k)=s(k+1)-s(k)$. It is clear that for Lyapunov function $V(k)=\frac{1}{2}s^{T}(k)s(k)>0$, $\Delta V(k)<0$ can be ensured by choosing   parameter $\bar{\xi} T>\bar{\delta}$ if $\parallel s(k)\parallel>\bar{\Phi}(k)$. Moreover, from $(\ref{UBS-27-1})$, the step-forward sliding mode function satisfy $\parallel s(k+1)\parallel\leq\sqrt{\parallel s(k)\parallel^{2}-2\eta(k)(\parallel s(k)\parallel-\bar{\Phi}(k))}$, and then we have $\parallel s(k+1)\parallel\leq\sqrt{\parallel s(k)\parallel^{2}+2\eta(k)\bar{\Phi}(k)}$. It can be concluded that (i): the sliding mode function will decrease when $\parallel s(k)\parallel>\bar{\Phi}(k)$ because $\Delta V(k)<0$; (ii): $s(k+1)$ satisfies
$\parallel s(k+1)\parallel\leq\sqrt{\bar{\Phi}^{2}(k)+2\eta(k)\bar{\Phi}(k)}$ when $\parallel s(k)\parallel\leq\bar{\Phi}(k)\leq\sqrt{\bar{\Phi}^{2}(k)+2\eta(k)\bar{\Phi}(k)}$. Therefore, the ultimate sliding  trajectories maintain in the QSMB (\ref{UBS-25}).

(2)
Note that the reachability of the system states can be guaranteed by selecting a large enough parameter $\bar{\xi} T$. Then, for sliding variable $G(k+1)$ solved by nonlinear matrix equations (\ref{UBS-21}), it must satisfy
\vspace{-0.1cm}
\begin{eqnarray*}
G(k+1)&=&[\bar{\mu}_{0}(k) s(k)+\bar{\xi} T\mathrm{sgn}s(k)-s(k)+\varpi_{1}(k-1)]\notag\\
&&\times L^{T}(k)\mathcal{X}^{T}_{1}(k)\{\mathcal{X}_{1}(k)L(k)L^{T}(k)\mathcal{X}_{1}(k)\}^{-1}.
\end{eqnarray*}
Denote
\begin{eqnarray*}
R_{g}\triangleq\left[
                                                 \begin{array}{cc}
                                                   \hat{R}_{g} & 1
                                                 \end{array}
                                               \right],~~
R_{1}\triangleq\left[
                                                 \begin{array}{cc}
                                                   \hat{R}_{1} & 0\\
                                                          0    & 1
                                                 \end{array}
                                               \right],
\end{eqnarray*}
with
\begin{eqnarray*}
\hat{R}_{g}&\triangleq&[\bar{\mu}_{0}(k) s(k)+\bar{\xi} T\mathrm{sgn}s(k)-s(k)+\varpi_{1}(k-1)]\notag\\
&&\times L^{T}(k)\mathcal{X}^{T}_{1}(k),\notag\\
\hat{R}_{1}&\triangleq&\mathcal{X}_{1}(k)L(k)L^{T}(k)\mathcal{X}_{1}(k).
\end{eqnarray*}
Apparently, $R_{g}$ is a matrix with appropriate dimension and $R_{1}>0$. And then we can conclude from Lemma \ref{lemma-1} that the SMD in (\ref{UBS-18-1}) will eventually stabilize for any $G(k+1)$ derived by (\ref{UBS-21}) if there exist scalar $\gamma > 0$ satisfy (\ref{UBS-8}) based on the above expression of $R_{g}$ and $R_{1}$. Moreover, the stability of the SMD (\ref{UBS-18-1}) can be guaranteed by the solved $G(k+1)$ due to $G(k+1)$ keeps the value calculated at the previous time unchanged when sliding mode surface enters the QSMB (\ref{UBS-25}).$\hfill\blacksquare$
\begin{remark}
For the solution of the equations (\ref{UBS-21}), denote $\omega(k)\triangleq \mathrm{col}\{L(k)$, $G^{(1)}(k+1)$, $\ldots,$ $G^{(n_{x}-n_{u})}(k+1), \bar{\mu}_{0}(k)\}$, where $G^{(i)}(k+1), i\in\{1,\ldots,n_{x}-n_{u}\}$ is the i-column of sliding variable $G(k+1)$. Let the initial  iteration value of vector parameter $\omega(k)$ is zero, the vector $\omega(k)$ can be derived through solving the nonlinear matrix equations (\ref{UBS-21}) by Levenberg-Marquardt algorithm, which can find the minimum value by gradient.
\end{remark}

Furthermore, the final implementation steps of the proposed input-mapping-based online learning SMC strategy are shown in Algorithm. \ref{alg-1}.


\begin{algorithm}
\caption{Co-design IMSMC strategy.}\label{alg-1}
Step 1.~Initialize $x(0)$ and $G(0)$ and assign the value to $\bar{\xi} T$. Set a value for $N$ and let $\mathcal{X}(0)=0$, $\mathcal{U}(0)=0$.\\
Step 2.~At the begin of $k$-th time, judge whether the condition $\parallel s(k)\parallel\leq\Omega$ is satisfied. If meet, jump to Step 3.  Otherwise, compute $L(k)$, $G(k+1)$ and $\bar{\mu}_{0}(k)$ according to (\ref{UBS-21}). Moreover, if the obtained $\bar{\mu}_{0}(k)\geq1$, set $\bar{\mu}_{0}(k)=0.99$. If  $\bar{\mu}_{0}(k)\leq0$, set $\bar{\mu}_{0}(k)=0.01$. Then, regain $L(k)$ and $G(k+1)$ by the equations composed of the first two expressions of (\ref{UBS-21}) based on the new $\bar{\mu}_{0}(k)$, and jump to Step 4.\\
Step 3.~Let $G(k+1)$ and $\bar{\mu}_{0}(k)$ keep the value calculated at the previous time, and solve $L(k)$ by the first relation of (\ref{UBS-21}).\\
Step 4.~Calculate remainder $\delta_{u}(k)$ of $u(k)$ by (\ref{UBS-15}) based on $\delta_{x}(k)=x(k)-\mathcal{X}(k-1)L(k)$ and the final value of $L(k)$, $G(k+1)$ and $\bar{\mu}_{0}(k)$. Then, $u(k)=\sum^{N}_{i=1}l_{i}u(k-i)+\delta_{u}(k)$ can be obtained. Implementing $u(k)$ on system (\ref{UBS-2}) and then measure the actual state $x(k+1)$.\\
Step 5.~Update the data $\mathcal{X}(k)$ and $\mathcal{U}(k)$ in memory by $x(k+1)$ and $u(k)$. Set $k=k+1$, and return to Step 2.
\end{algorithm}
\section{Illustrative Example}
For demonstrating the validity and effectiveness of the proposed input-mapping-based online learning SMC strategy for considered system (\ref{UBS-2}), some simulations are displayed in this section.
Firstly, consider a numerical example with the following parameters
\allowdisplaybreaks
\begin{eqnarray*}
A&=&\left[
    \begin{array}{ccc}
      0.1012   & 0.8075  &  1.7837\\
   -0.0529   & 0.0944  & -0.0396\\
         0    &0.1937   & 0.5402
    \end{array}
  \right],~~B=\left[
                \begin{array}{c}
                  0\\
                  0\\
                  0.1
                \end{array}
              \right],\\
\bar{D}&=&\left[
                \begin{array}{ccc}
                  0.2&
                  0.1&
                  0.2
                \end{array}
              \right]^{T},~~\bar{E}=\left[
                \begin{array}{ccc}
                  0.5&
                  0.2&
                  0.1
                \end{array}
              \right].
\end{eqnarray*}
Then, based on the above system parameters, the following two examples are given to make some comparisons for the proposed algorithm with the robust method in \cite{QXZ2014TIE} and the data-driven method in  \cite{MXLZ2020TCS,EOR2018IJC}.
\subsection{Example 1}
In this section, an example is given to compare the proposed co-design IMSMC method, the robust method proposed in \cite{QXZ2014TIE} and the adaptive method in \cite{MXLZ2020TCS} . By solving LMI (\ref{UBS-8}) with the above system parameters, the sliding mode variable $G$ for the robust SMC method is obtained as $G=[0.0728~~0.4562]$. Other parameters in \cite{QXZ2014TIE} are given as $\bar{\mu}_{0}=0.1$,  $\bar{\xi} T=0.01$. For a fair comparison, the sliding mode variable $G$ in \cite{MXLZ2020TCS} is set as robust method, and the remaining parameters are $\delta_{0}=150$, $Tp_{0}=Tp_{1}=0.0004$, $\alpha_{0}=\alpha_{1}=1$, $q_{0}=0.5$, $T\hat{k}_{0}(0)=0.01$ and $T\hat{k}_{1}(0)=0$. For the proposed method in this paper, the length of the historical data is chosen as $N=2$, and the switching parameter is set as the robust method. In addition, a disturbance $f(k)=[0~0~1]^{T}$ is applied to the real system at $k=50$ to $k=95$  for further comparison. For initial state $x(0)=[-1~~1~-5]^{T}$, the effectiveness and superiority of the proposed method are presented in the following two cases.

\begin{figure*}[h]
\centering
\subfigure[State $x_{1}(k)$.]{
\includegraphics[width=6.1cm]{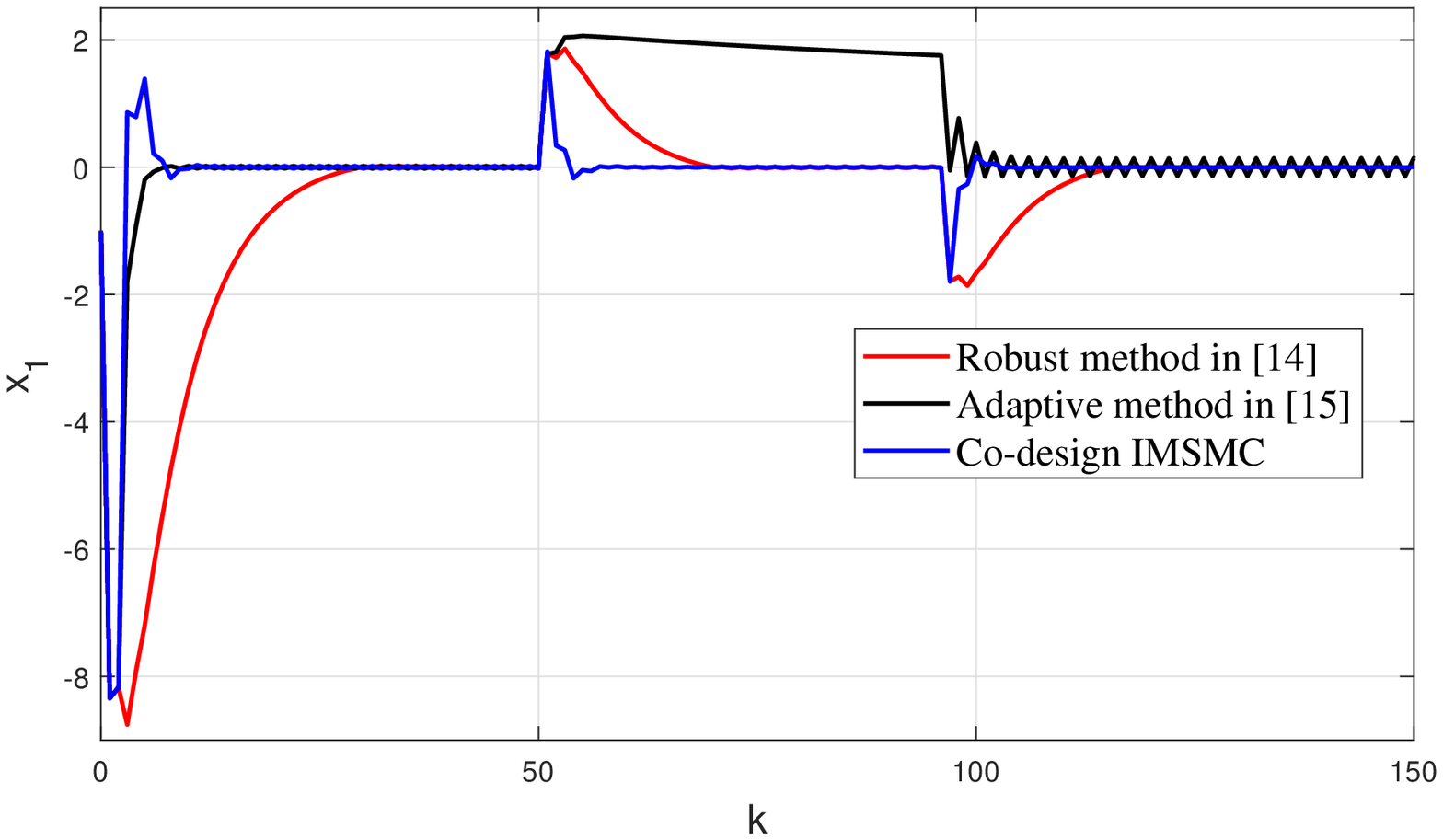}
\label{Fig-1-1-1}
}
\hspace{-8mm}
\subfigure[State $x_{2}(k)$.]{
\includegraphics[width=6.1cm]{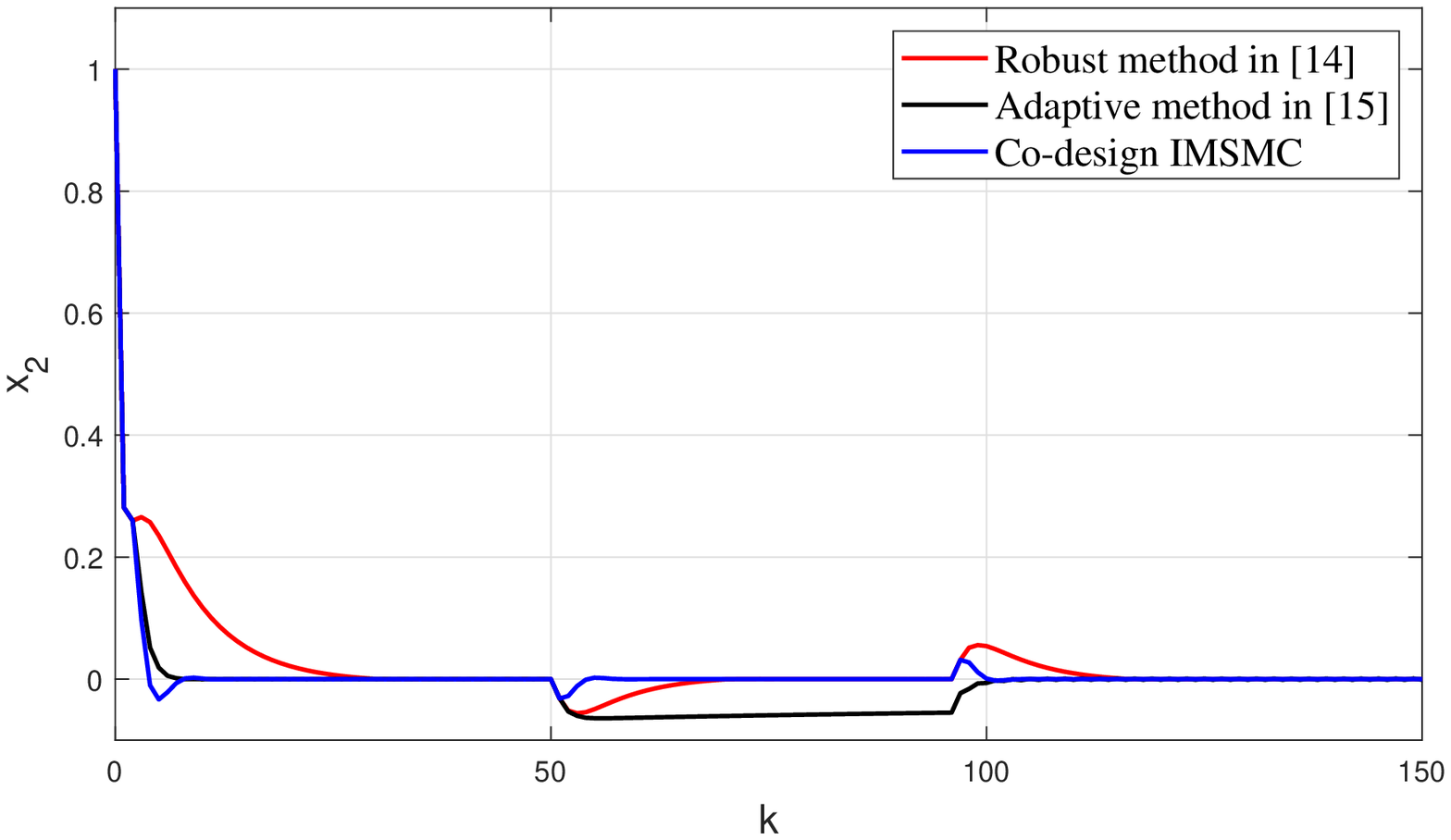}
\label{Fig-1-1-2}
}
\hspace{-8mm}
\subfigure[State $x_{3}(k)$.]{
\includegraphics[width=6.1cm]{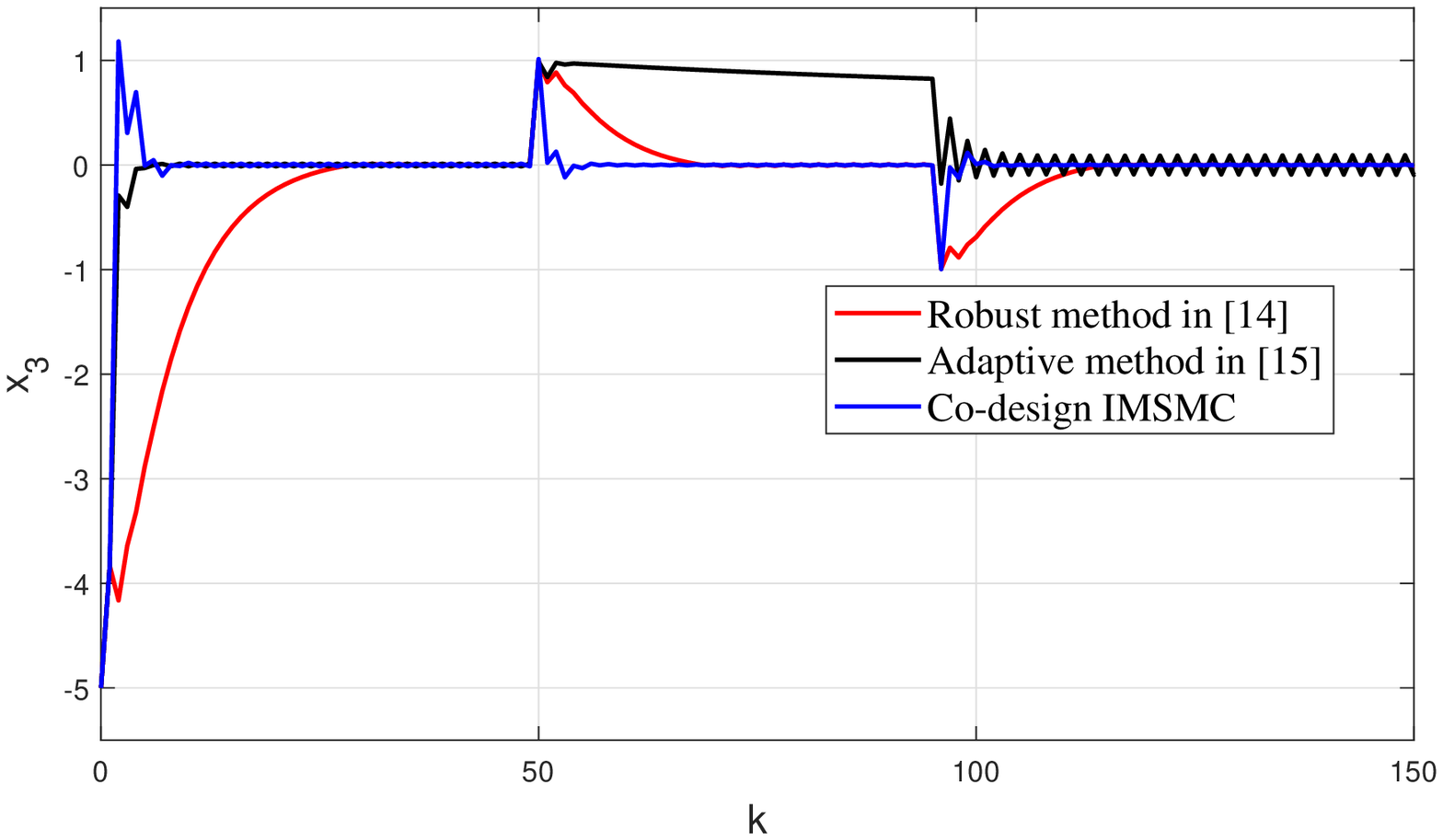}
\label{Fig-1-1-3}
}
\hspace{-8mm}
\subfigure[Control input $u(k)$.]{
\includegraphics[width=6.1cm]{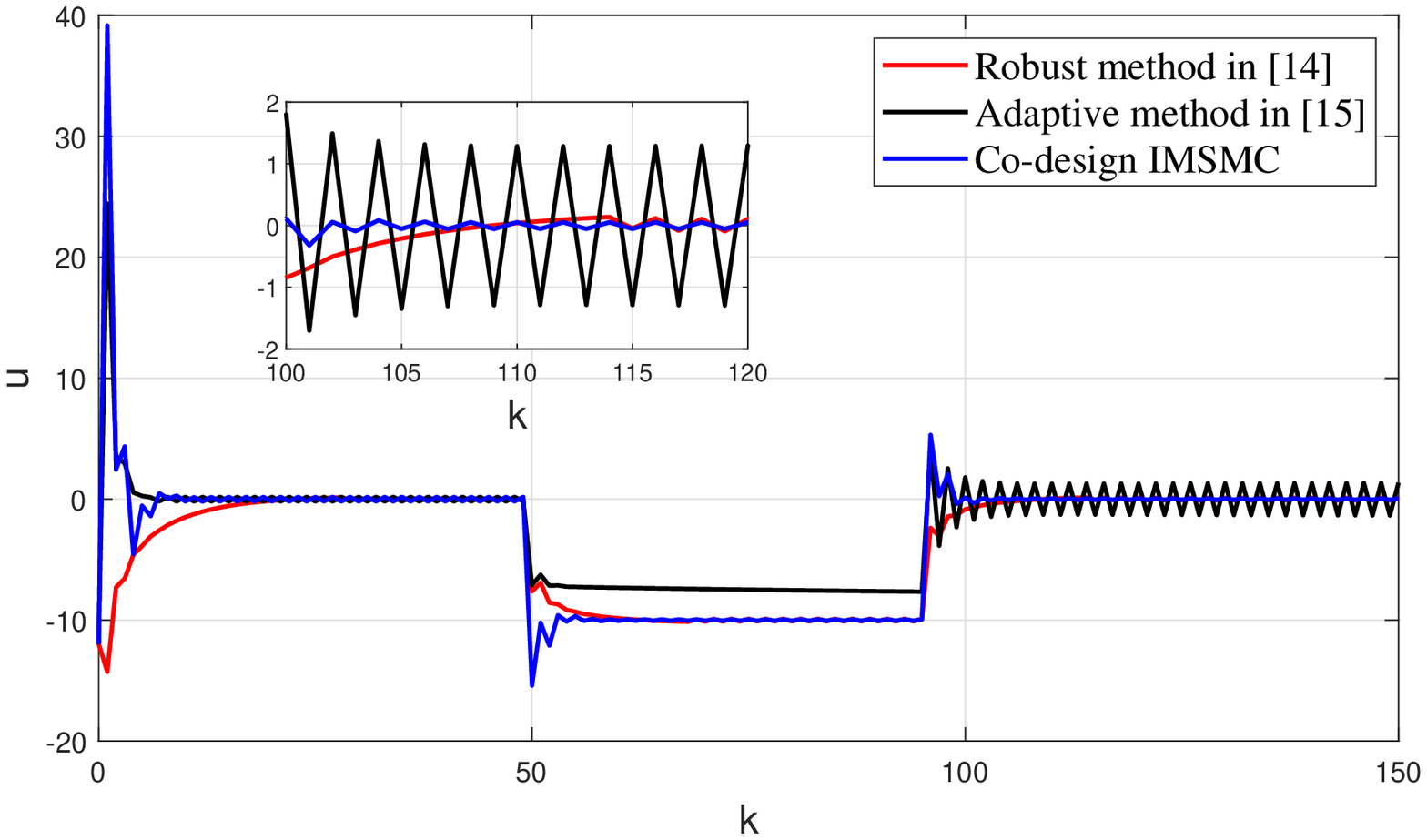}
\label{Fig-1-1-4}
}
\hspace{-8mm}
\subfigure[Sliding mode surface $s(k)$.]{
\includegraphics[width=6.1cm]{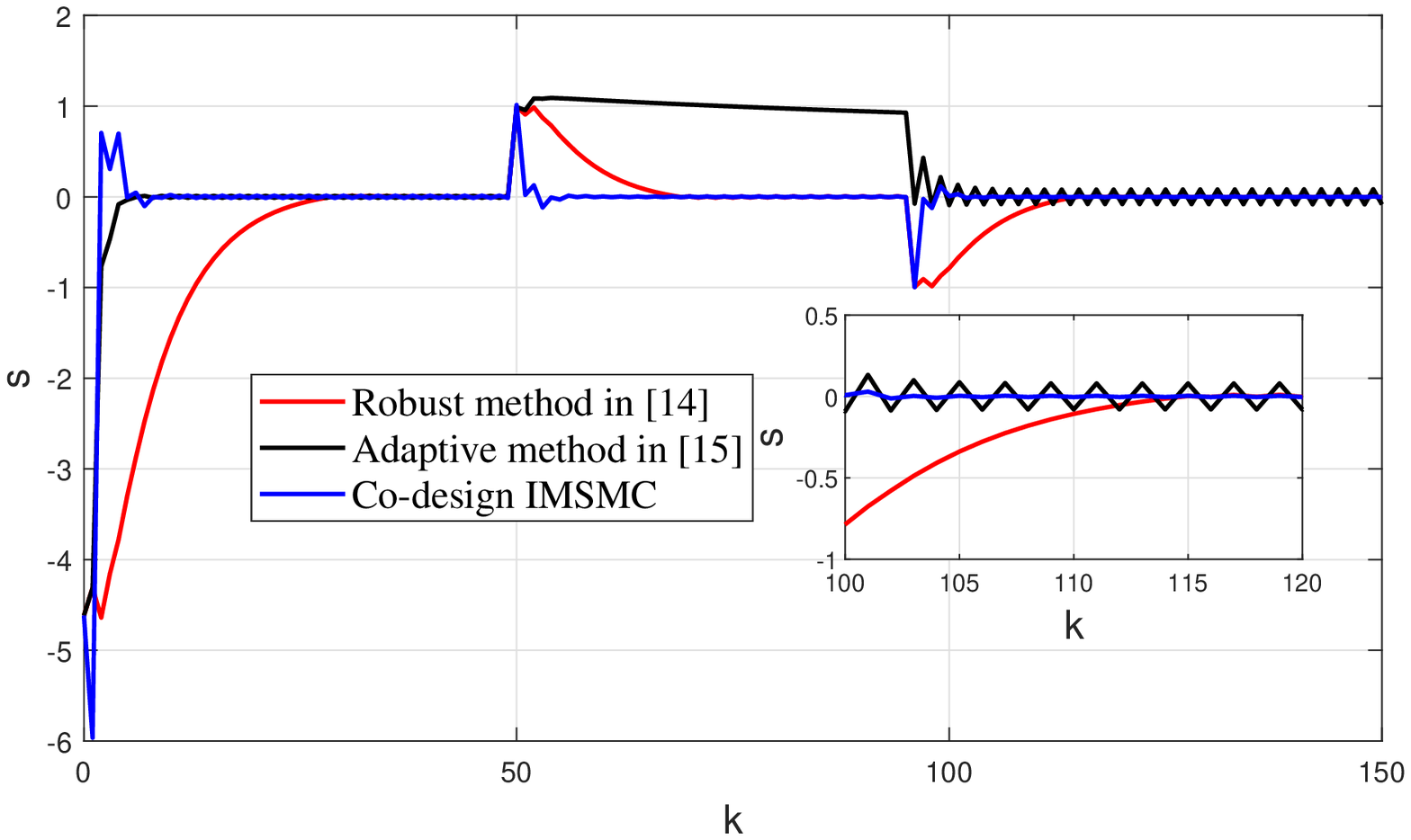}
\label{Fig-1-1-5}
}
\caption{The curves when $\Delta=0.8$.}
\label{Fig-1-1}
\end{figure*}

Case 1:~Suppose that $\Delta=0.8$, the response curve of the closed-loop system is depicted in Fig. \ref{Fig-1-1}. It can be observed from Figs. \ref{Fig-1-1-1}-\ref{Fig-1-1-3} that the convergence rate of system states in terms of control law (\ref{UBS-16}) and the adaptive control law in \cite{MXLZ2020TCS} are faster than that of robust SMC law. For the adaptive in \cite{MXLZ2020TCS}, the parameters needed to be tuned by trial-and-error method to obtain a satisfactory performance, which reduces its application efficiency. And the system state is difficult to reach the equilibrium point if the system is disturbed during operation. Figs. \ref{Fig-1-1-4}-\ref{Fig-1-1-5} show the corresponding control law and sliding mode surface, from which we have the proposed co-design IMSMC can not only keep a small chattering and improve the performance of the system, but also better overcome the impact of the burst disturbance.

\begin{figure*}[h]
\centering
\subfigure[State $x_{1}(k)$.]{
\includegraphics[width=6.1cm]{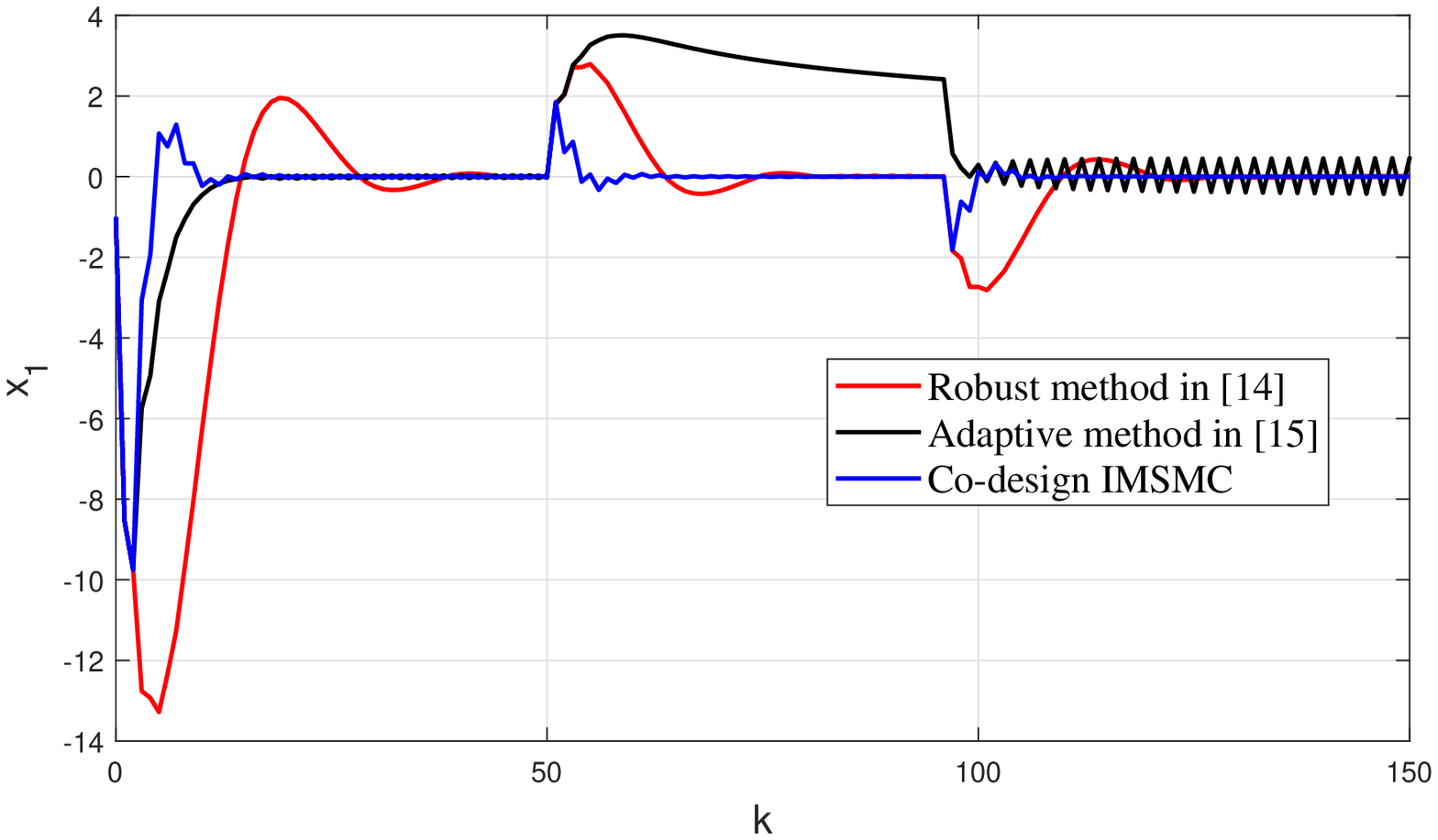}
\label{Fig-2-1-1}
}
\hspace{-8mm}
\subfigure[State $x_{2}(k)$.]{
\includegraphics[width=6.1cm]{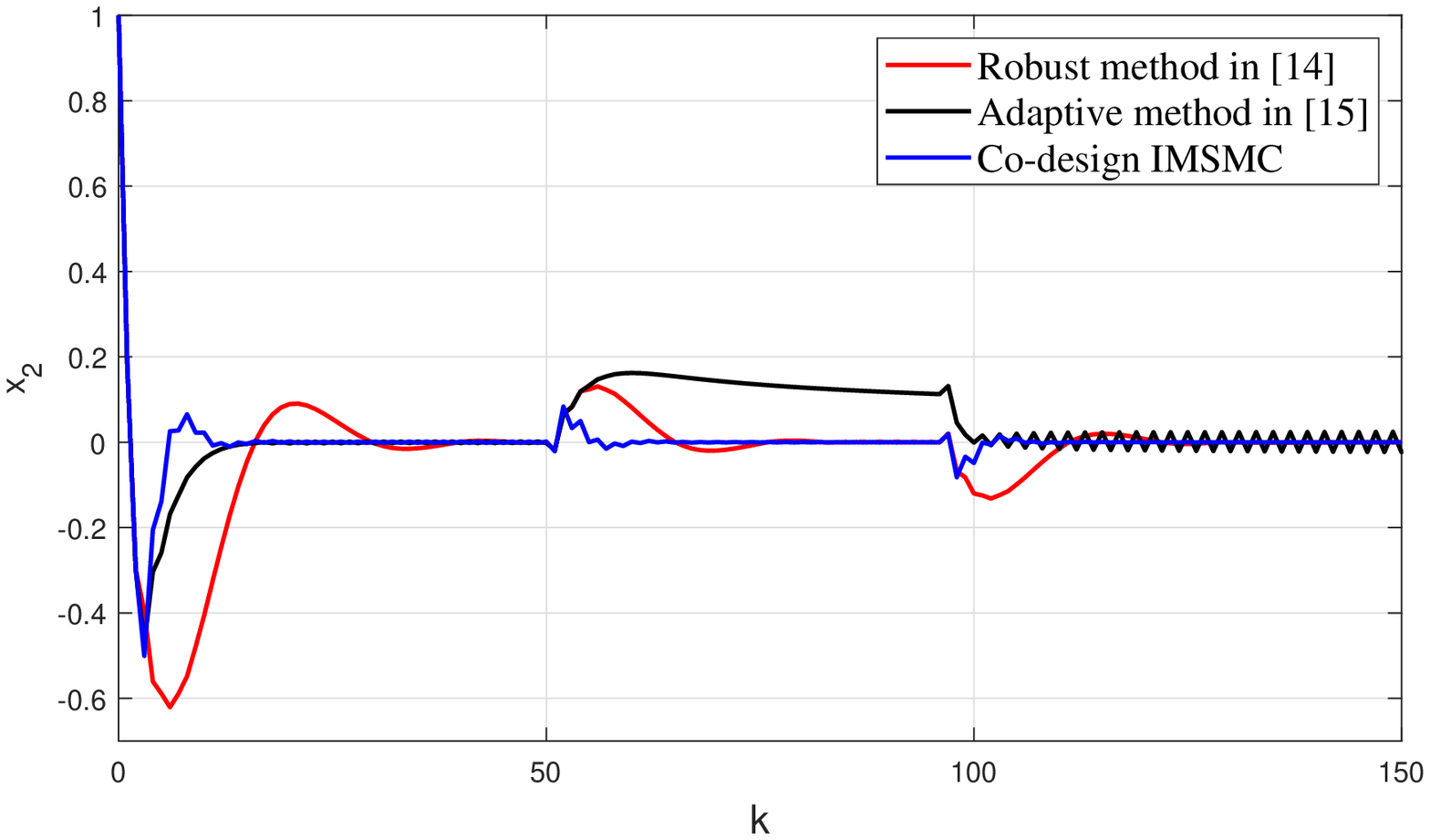}
\label{Fig-2-1-2}
}
\hspace{-8mm}
\subfigure[State $x_{3}(k)$.]{
\includegraphics[width=6.1cm]{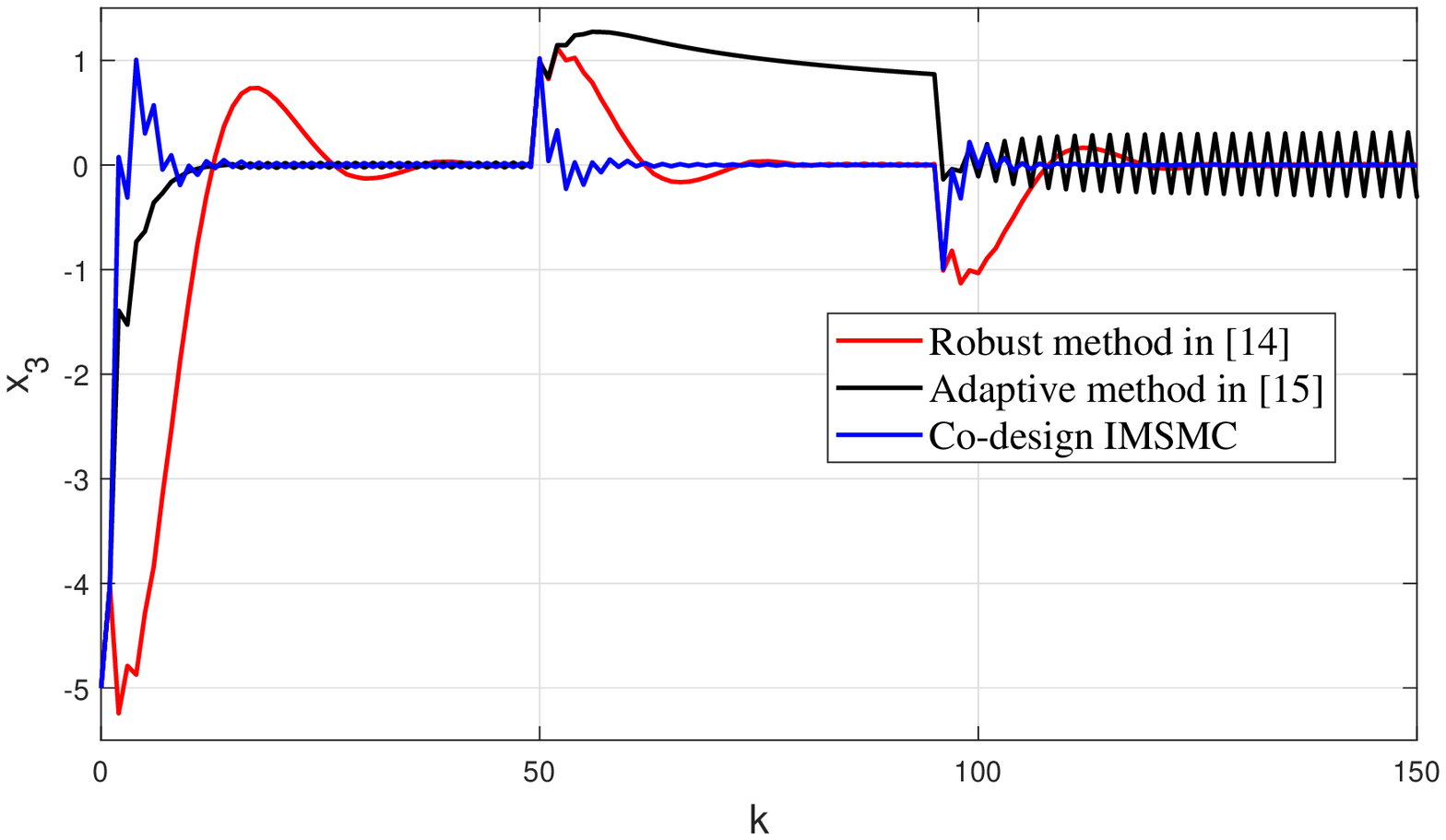}
\label{Fig-2-1-3}
}
\hspace{-8mm}
\subfigure[Control input $u(k)$.]{
\includegraphics[width=6.1cm]{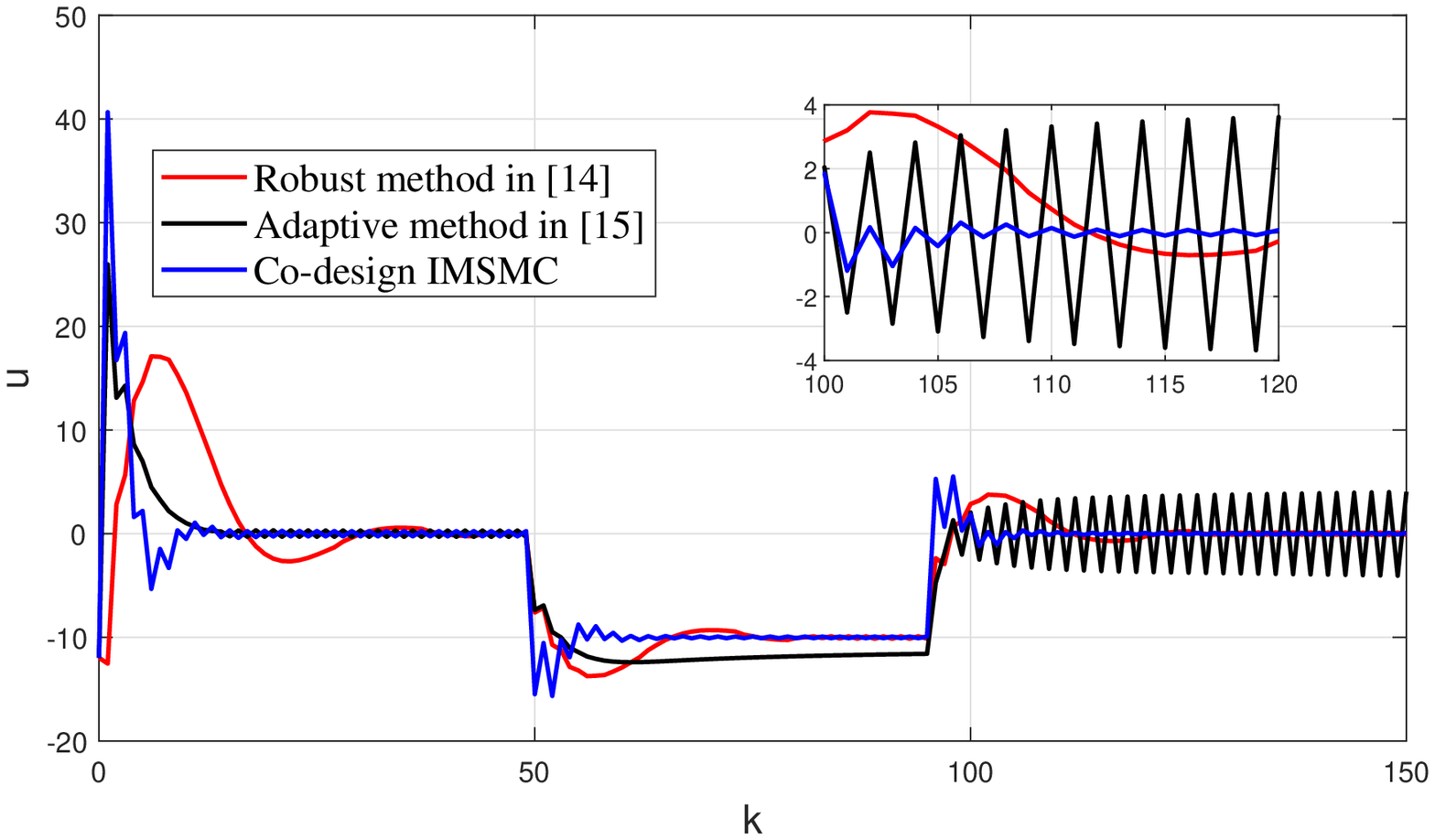}
\label{Fig-2-1-4}
}
\hspace{-8mm}
\subfigure[Sliding mode surface $s(k)$.]{
\includegraphics[width=6.1cm]{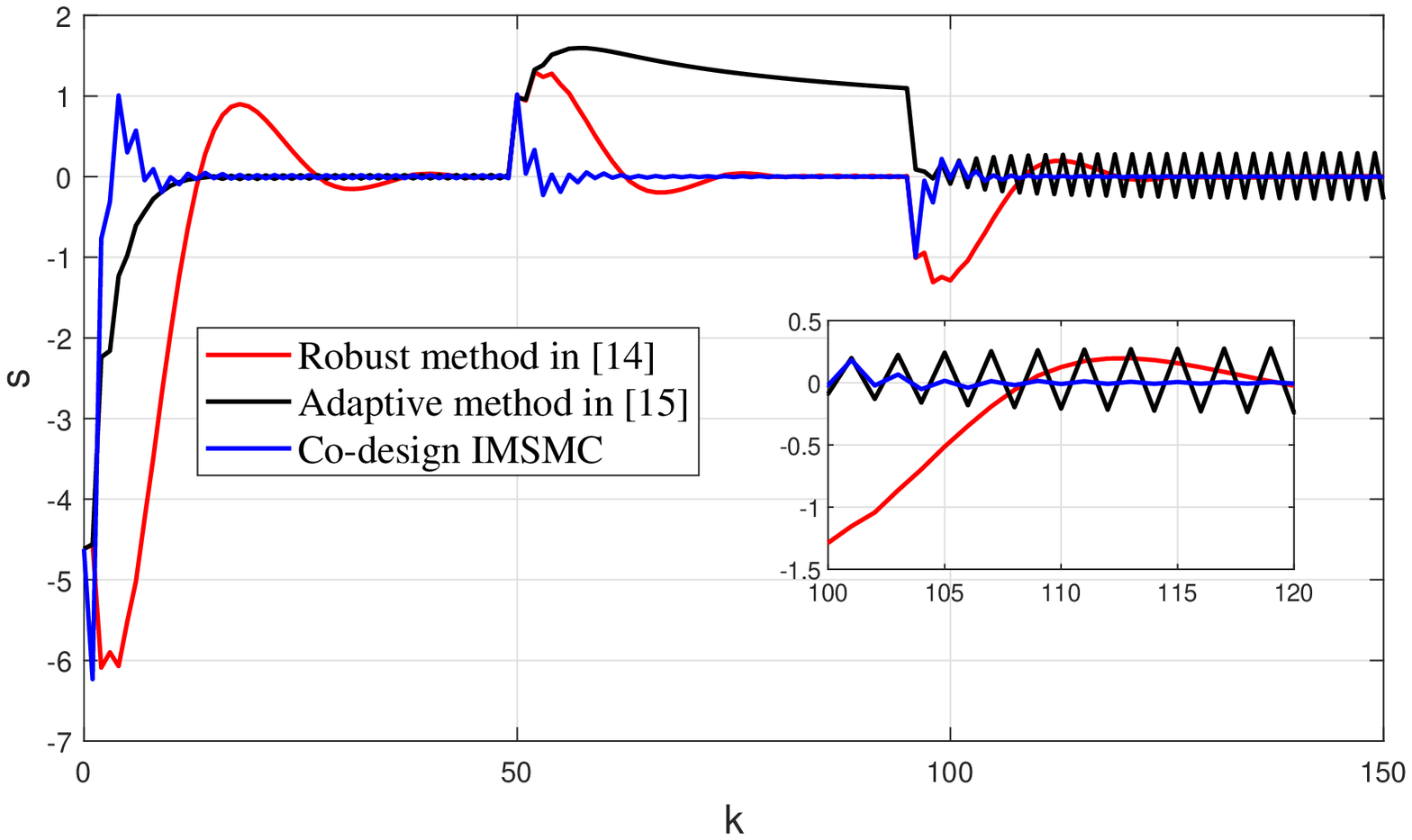}
\label{Fig-2-1-5}
}
\caption{The curves when $\Delta=2$.}
\label{Fig-2-1}
\end{figure*}

Case 2:~Suppose that $\Delta=2$, which is beyond the constraints of unknown dynamic. The corresponding response curves of the closed-loop system are given in Figs. \ref{Fig-2-1-1}-\ref{Fig-2-1-3}. There is no doubt that the improvement of the convergence rate by using the co-design IMSMC method in this paper is obvious. Furthermore, the SMC law and sliding mode surface are shown in Figs. \ref{Fig-2-1-4}-\ref{Fig-2-1-5}. It is easy to see that the robustness of the proposed co-design IMSMC strategy is stronger.

\subsection{Example 2}
\begin{figure*}[!htp]
\centering
\subfigure[System output $y(k)$.]{
\includegraphics[width=6.1cm]{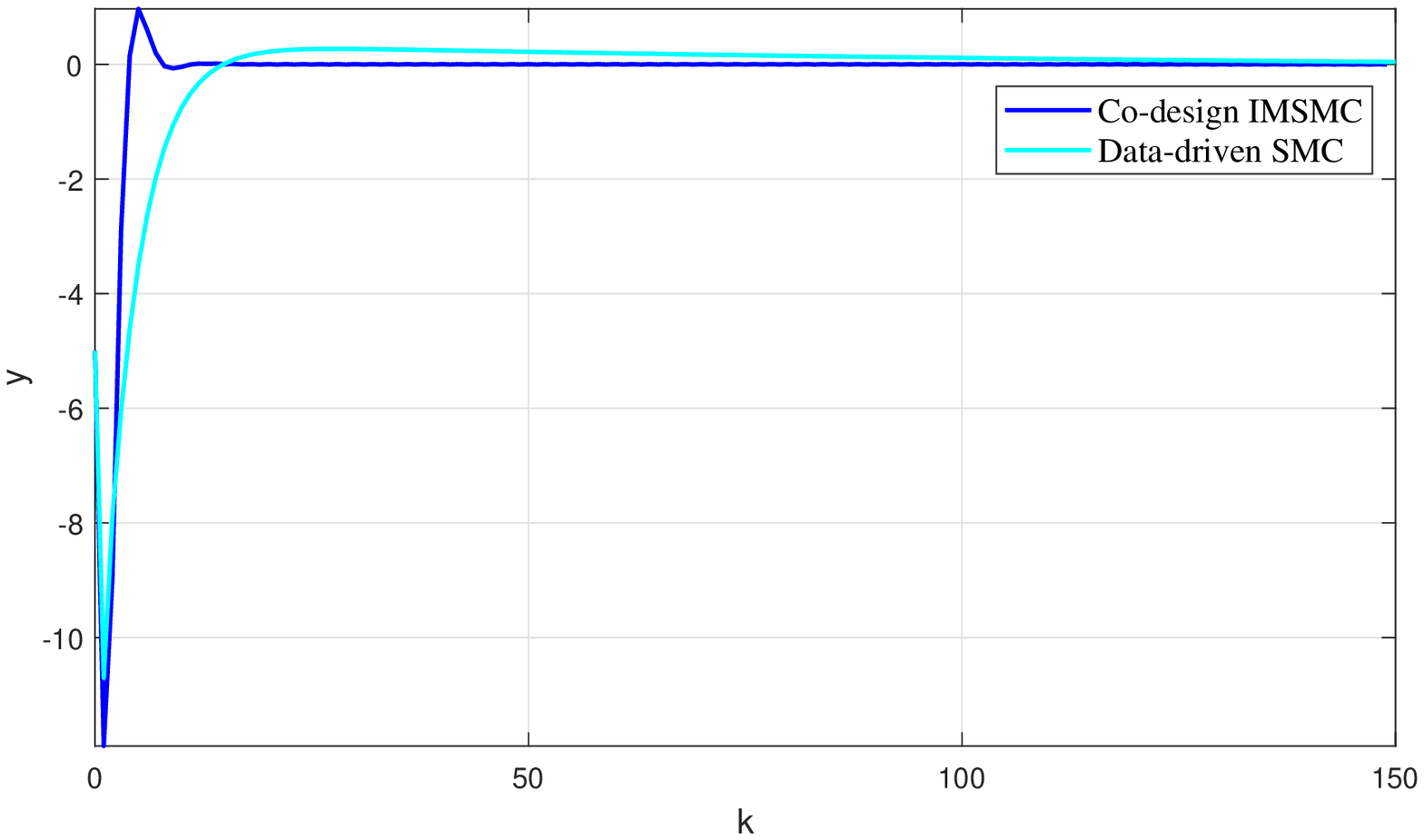}
}
\hspace{-8mm}
\subfigure[Control input $u(k)$.]{
\includegraphics[width=6.1cm]{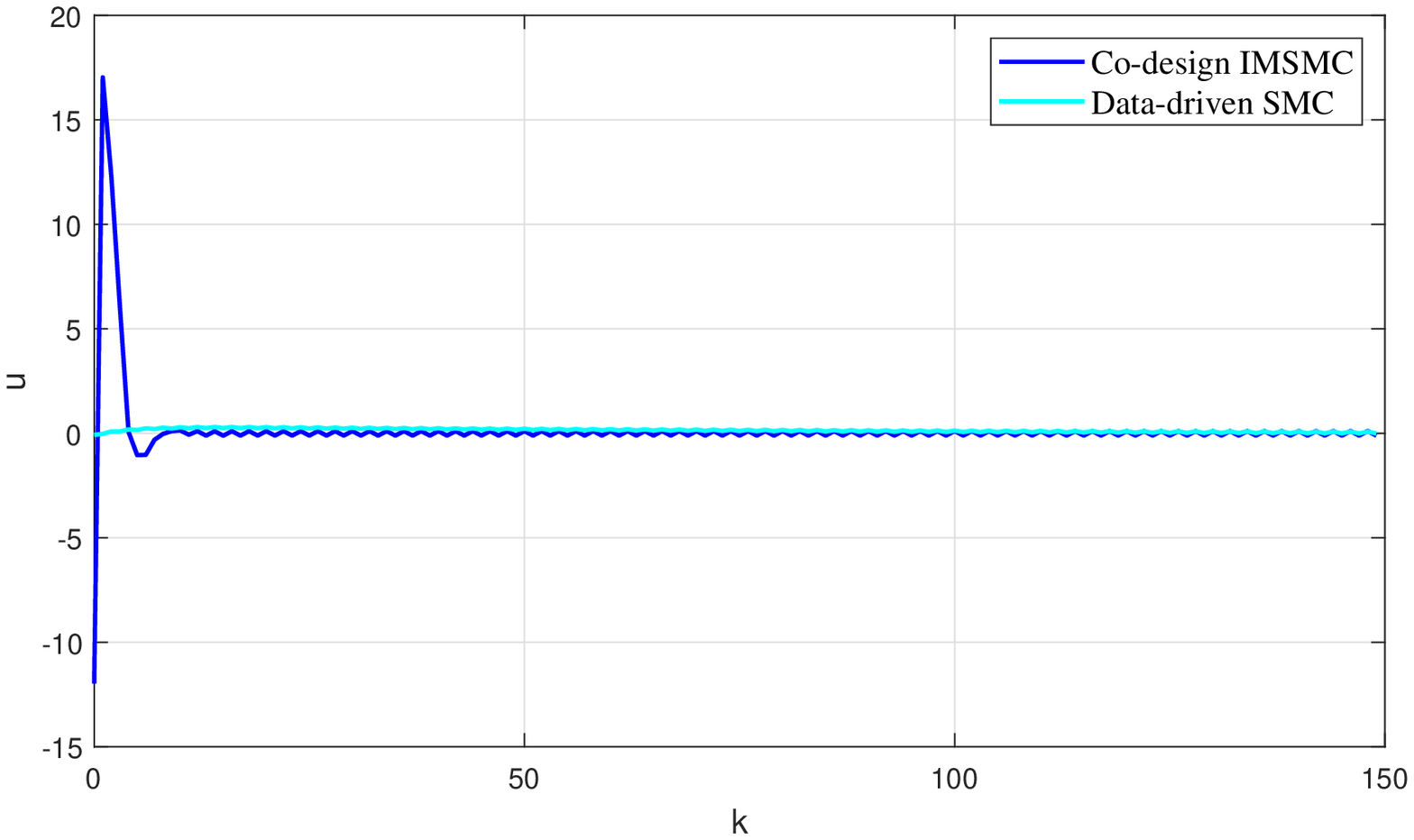}
}
\hspace{-8mm}
\subfigure[Sliding mode surface $s(k)$.]{
\includegraphics[width=6.1cm]{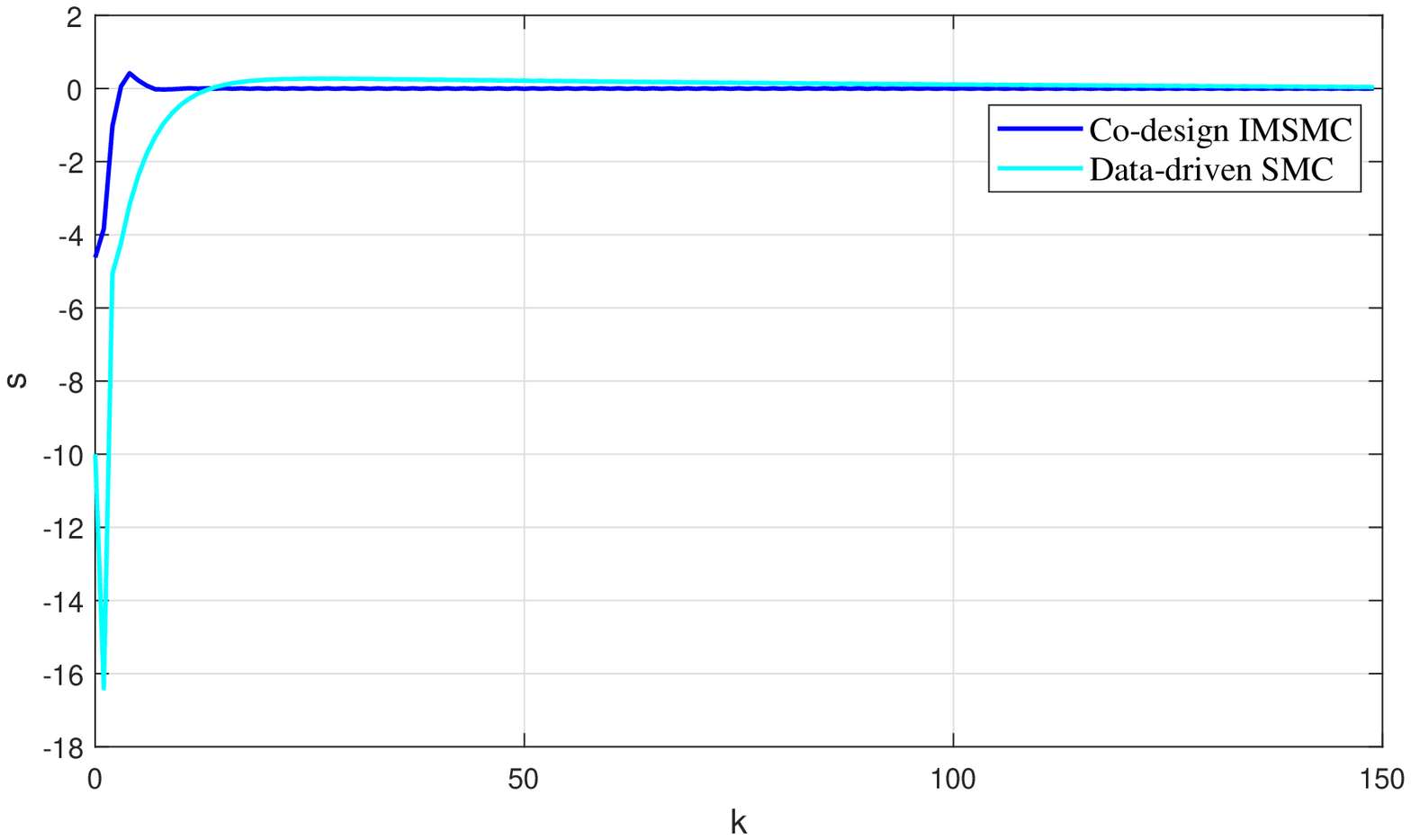}
}
\caption{The comparison of data-driven SMC and co-design IMSMC.}
\label{Fig-3-1}
\end{figure*}

 Now, the proposed co-design IMSMC law is compared with the data-driven SMC method in \cite{EOR2018IJC}, which uses an identification method to identify the model. Firstly, the simulation result of the paper is reproduced for a correct comparison. Secondly, the output of the system (\ref{UBS-3}) and the tracking object are supposed to be $y(k)=\left[\begin{array}{ccc}1 & 1 & 1 \\ \end{array} \right]x(k)$ and  $y_{d}(k)=0$, respectively.  Then, the parameters in \cite{EOR2018IJC} are set as $\mu=1.2, \eta=0.8, \beta=9, \lambda=1, \alpha=0.1, n=1, \delta=0.01, \gamma=0.5$. However, the system output $y(k)$ is always diverging until the initial value of the estimated pseudo gradient $\hat{\emptyset}(0)$ reaches  to $\hat{\emptyset}(0)=40$. Next, the parameters of the proposed co-design IMSMC method are chosen as Example 1. It can be observed from the simulation result in Figure. \ref{Fig-3-1} that the proposed method can make the convergence rate of the considered system faster. Thus, we can come to a conclusion that the model-free adaptive method in \cite{EOR2018IJC} requires identifying a model, and the identification results have high requirements on the initial value of the model parameter. However, the co-design IMSMC method proposed in this paper is unnecessary to identify model parameters, and the future system dynamic is directly predicted based on the historical data. Furthermore, the convergence rate of the addressed system is improved more significantly by the results in this paper.

\section{Conclusion}
A novel online learning SMC strategy for the system is studied in this paper via an input-mapping method. Firstly, a new form of the model is established based on the input-mapping of historical information and a part of the model is known accurately which is beneficial for improving the system performance. Secondly, the input-mapping technique is introduced detailed to further explain the motivation and principle of the method. Then, the sliding mode surface and the IMSMC law are co-designed by minimizing an optimization objective, and the combination coefficients of the historical information are also derived through optimization. Next, the whole algorithm of the proposed co-design IMSMC strategy and the stability analysis are presented. Finally, the effectiveness of the proposed method is illustrated by a simulated system and  some given comparisons.
\bibliographystyle{elsarticle-num}
\bibliography{myreferences}

\end{document}